\documentclass{article}
\usepackage{amsmath}
\usepackage{amssymb}
\usepackage{graphicx}
\usepackage{hyperref}
\usepackage{geometry}
\usepackage{color}
\usepackage{tabularray}
\usepackage{array}
\usepackage{bigstrut}
\usepackage{multirow}
\usepackage{bm}
\usepackage{booktabs}
\usepackage[numbers]{natbib}

\begin{document}

\title{Enhancing Context Models for Point Cloud Geometry Compression with Context Feature Residuals and Multi-Loss}
\author{Chang Sun, Hui Yuan, Shuai Li, Xin Lu, and Raouf Hamzaoui}
\date{}
\maketitle

\begin{abstract}
In point cloud geometry compression, context models usually use the one-hot encoding of node occupancy as the label, and the cross-entropy between the one-hot encoding and the probability distribution predicted by the context model as the loss function. However, this approach has two main weaknesses. First, the differences between contexts of different nodes are not significant, making it difficult for the context model to accurately predict the probability distribution of node occupancy. Second, as the one-hot encoding is not the actual probability distribution of node occupancy, the cross-entropy loss function is inaccurate. To address these problems, we propose a general structure that can enhance existing context models. We introduce the context feature residuals into the context model to amplify the differences between contexts. We also add a multi-layer perception branch, that uses the mean squared error between its output and node occupancy as a loss function to provide accurate gradients in backpropagation. We validate our method by showing that it can improve the performance of an octree-based model (OctAttention) and a voxel-based model (VoxelDNN) on the object point cloud datasets MPEG 8i and MVUB, as well as the LiDAR point cloud dataset SemanticKITTI. 
\end{abstract}

\section{Introduction}
Point clouds are representations of three-dimensional (3D) objects or scenes as a set of points. Point clouds have a wide range of applications in immersive real-time communication, virtual reality, automated driving \cite{1}, etc. The points forming a point cloud are defined by their geometry information (Cartesian coordinates $x,y,z$) and attribute information (e.g., color or reflectance). Due to the huge amount of data contained in a point cloud, efficient compression is necessary to reduce storage and transmission costs. However, unlike images and videos, which are regularly distributed in 2D space, point clouds are sparse and irregularly distributed in 3D space, making efficient compression more challenging.

The geometry-based point cloud compression (G-PCC) \cite{2} standard proposed by Moving Picture Experts Group (MPEG) uses either octree coding \cite{3}\cite{4} or a pruned octree combined with the triangle soup technique \cite{5}\cite{6} to encode the geometry information. The triangle soup method has better rate-distortion performance. However, unlike the octree coding method, it is not lossless. The octree coding method recursively decomposes the point cloud geometry information and encodes the occupancy of the octree nodes at each level using an adaptive arithmetic coder based on a hand-crafted context model \cite{7,8,9,10}. Specifically, the probability distribution of the node occupancy is predicted by the context model and guides the encoding for the arithmetic encoder.

Recently, learning-based context models have been used to improve the efficiency of point cloud compression. Huang et al. \cite{11} proposed an octree-based context model called OctSqueeze, which has shown superior performance compared to hand-crafted context models. OctSqueeze takes the ancestor nodes as context and uses a multi-layer perception (MLP) network to output the probability distribution of the octant occupancy.  Several works \cite{12,13,14,15} have improved the performance of OctSqueeze by introducing more context information. VoxelDNN \cite{16}, a voxel-based context model, uses a deep convolutional neural network with masked filters to learn a probability distribution of the voxels. To the best of our knowledge, most context models for point cloud geometry compression use decoded sibling nodes as context. Furthermore, most octree-based context models use a 255-dimensional one-hot encoding based on the occupancy of the child node as the label during training and the cross-entropy between the one-hot encoding and the probability distribution predicted by the context model as the loss function. This approach essentially transforms the prediction problem of node occupancy probability distribution into a 255-dimensional classification problem, leading to two issues:
   
\begin{figure}[!t]
\centering
\includegraphics[width=8cm]{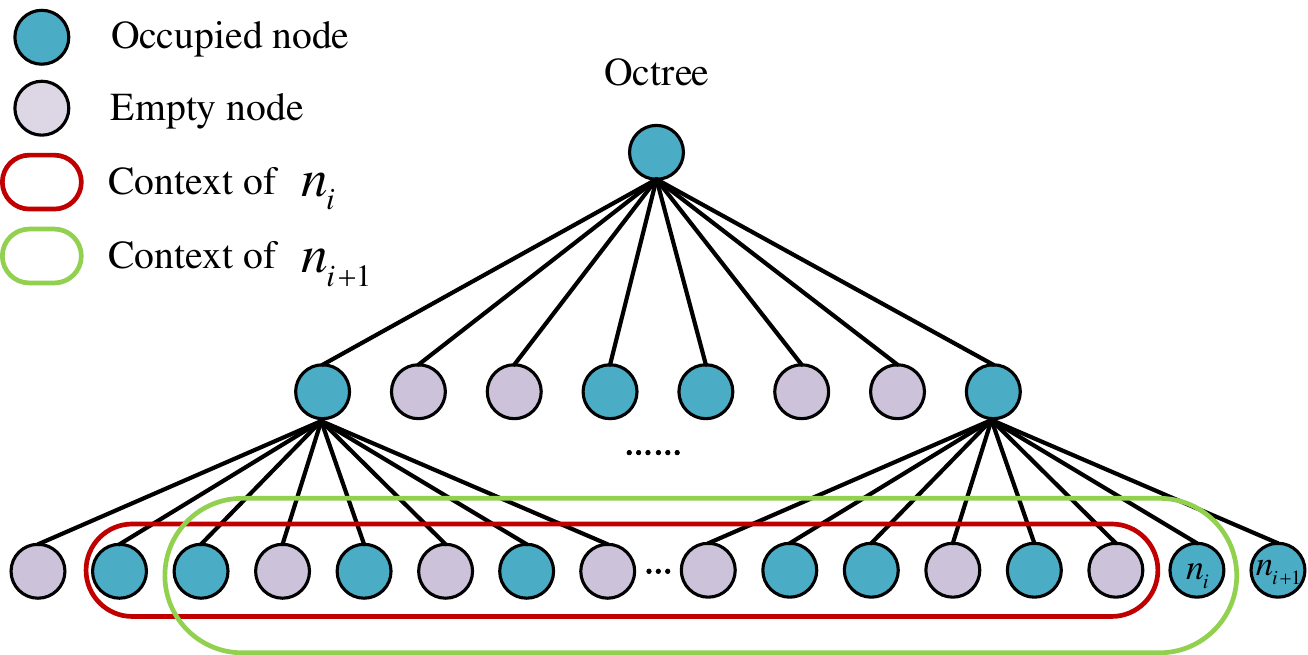}
\caption{Differences in context between adjacent nodes. Context models typically use decoded nodes as context, resulting in small differences in context between neighboring nodes.}
\label{fig1}
\end{figure}  
  
\begin{enumerate}
\item {The differences in context between different nodes are not significant enough for a context model to accurately predict the probability distribution of node occupancy. In Fig. 1, for example, the contexts of two adjacent nodes differ by one node only. Therefore, it is required to amplify the differences between contexts of different nodes.}
\item {The one-hot encoding of node occupancy does not reflect the true probability distribution of node occupancy, leading to inaccurate learning goals of the context model. Moreover, as shown in Fig. 2, the cross-entropy fails to accurately measure the differences between one-hot encoding and probability distribution. Therefore, we need to introduce accurate label and loss functions for effective training.}
\end{enumerate}

Instead of proposing a special network, we propose a general structure that can enhance existing context models. The contributions of our work can be summarized as follows:
\begin{enumerate}
\item {To enhance the differences between contexts, we propose to include context feature residuals of adjacent contexts into the context models. Furthermore, we use the cosine similarity and the Euclidean distance to calculate the inter-class differences in context. Our results demonstrate that using context feature residuals significantly amplifies inter-class differences.}
\item {We improve the performance of the context models by adding an MLP branch that directly predicts the node occupancy instead of the probability distribution. The loss function of this branch is the mean squared error (MSE) between its output and the actual node occupancy. Since the node occupancy is an accurate label, this branch introduces accurate gradients during the training of the context model. At the same time, the output of this branch will also serve as a feature to assist the training of the main network.}
\item {We demonstrate the effectiveness of our approach by applying it to two state-of-the-art models: an octree-based one (OctAttention \cite{14}) and a voxel-based one (VoxelDNN \cite{16}). Experimental results on object point cloud datasets MPEG 8i and MVUB, as well as LiDAR point cloud dataset SemanticKITTI show that our method can reduce the bitrate in geometry point cloud encoding without significantly increasing time complexity.}
\end{enumerate}

The remainder of this paper is organized as follows. Section 2 reviews context models for 3D point cloud geometry coding. Section 3 describes the proposed method. Section 4 presents the experimental results and analysis. Finally, Section 5 gives our conclusions and suggests future work.

\begin{figure}[!t]
\centering
\includegraphics[width=8cm]{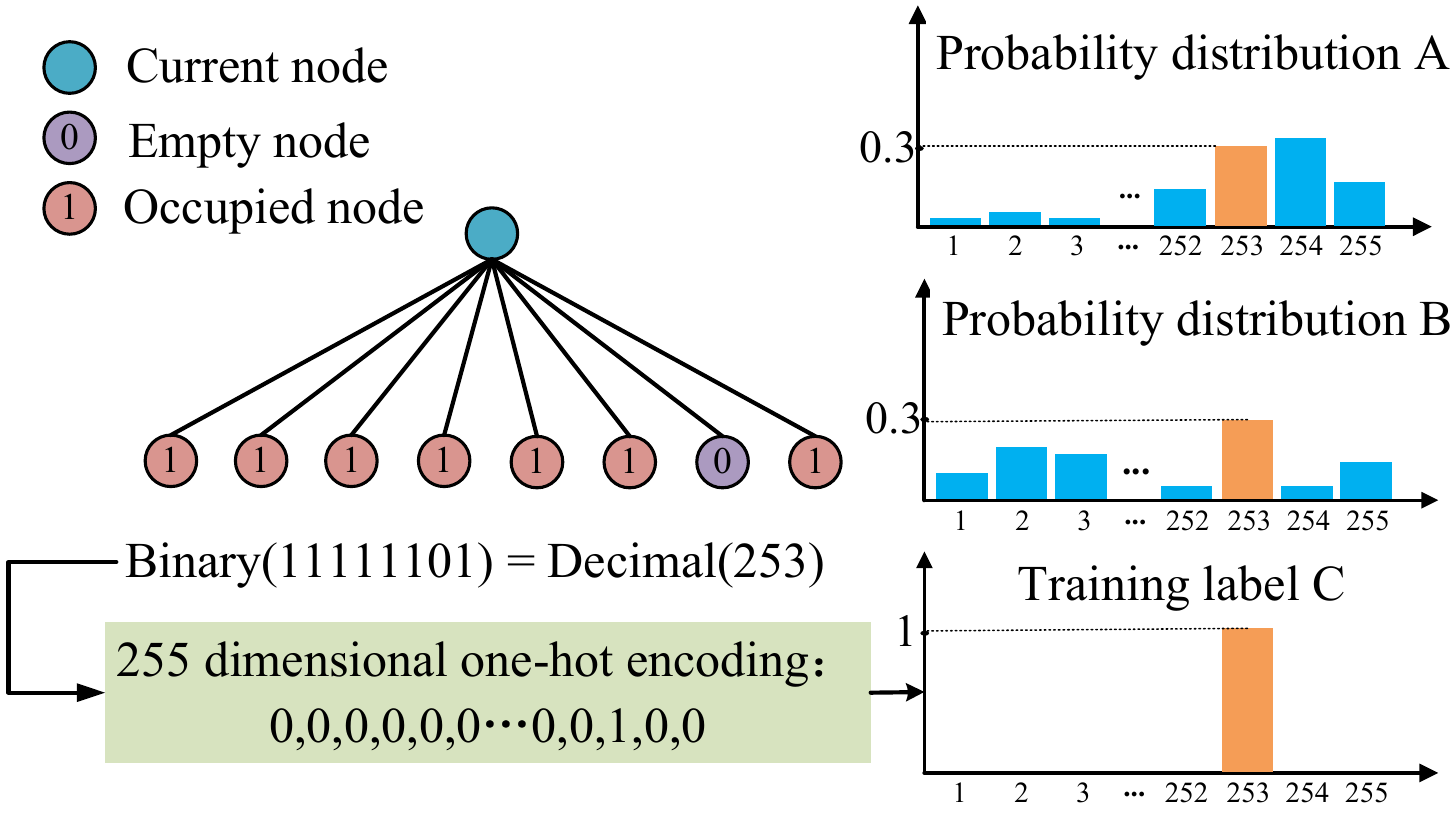}
\caption{Example of one-hot encoding based cross-entropy loss. Excluding the case where all eight child nodes are empty, there are 255 possible occupancy configurations of the eight child nodes, which can be represented using a 255-dimensional one-hot encoding. The 255-dimensional one-hot encoding is commonly used as a training label for octree-based context models.}
\label{fig2}
\end{figure}

\section{Related Work}
The G-PCC standard \cite{17} proposed by MPEG is commonly used for compressing object point clouds and LiDAR point clouds. Octree analysis in G-PCC recursively decomposes the point cloud geometry information using an octree. Each node in the octree is encoded by an adaptive arithmetic coder based on a hand-crafted context model. According to Shannon’s source coding theorem, the probability distribution predicted by the context model significantly affects the efficiency of arithmetic encoding. Deep learning methods \cite{18,19,20,21} have gained increasing popularity in the field of point cloud compression.  Learning-based context models have shown superior performance compared to hand-crafted context models.

Learning-based context models can be used for end-to-end cloud compression. The works in \cite{22,23,24,25,26,27} use neural networks to obtain the latent representation of point clouds. Subsequently, the context model predicts the probability distribution of the latent representation and further enhances compression efficiency through entropy coding. Although the entropy coding of the latent representation is lossless, mapping the latent representation to point clouds through the learning based model introduces distortion. 
 
 Learning-based context models can also be used for lossless compression. Huang et al. \cite{11} propose a learning-based context model called OctSqueeze that takes ancestor nodes as context and outputs a probability distribution of node occupancy through an MLP network. OctSqueeze outperforms hand-crafted context models, but both the context which is composed solely of ancestor nodes and the model which consists solely of an MLP are too simplistic. MuSCLE \cite{12} introduces an inter-frame context which is composed of ancestor nodes and neighboring nodes at the same hierarchical level in the frame that has been encoded. However, this work ignores the neighboring nodes in the current point cloud frame and local geometry features such as quadratic surfaces. Chen et al. \cite{15} introduces a neural network to locally fit quadratic surfaces. The features of a quadratic surface are combined with ancestor nodes, neighboring nodes, and sibling nodes to form the context. VoxelContext-Net \cite{13} constructs a context based on the sibling and neighboring nodes, but the context is limited to $9\times9\times9$ nodes and the receptive field of the context becomes smaller as the encoding node resolution increases. OctAttention \cite{14} introduces a large-scale context, including ancestor nodes, thousands of neighboring nodes, and ancestor nodes of these neighboring nodes. The context almost covers the information of the entire octree. However, such heavy global attention computations and auto-regressive contexts are inefficient for practical applications. Recently, Nguyen and Kaup \cite{28} proposed CNet, which uses sparse tensors to represent all encoded points to construct the context and predicts the occupancy probability and attribute probability of the leaf node through a sparse convolutional network. The method shows outstanding performance in geometry and attribute lossless compression. However, using decoded sibling nodes as context leads to a slow decoding process. Similar to the L3C \cite{29} image compression method, some methods generate point clouds of various resolutions via down-sampling and predict the probability distribution in a coarse-to-fine order to improve the decoding speed. MS-VoxelDNN \cite{30} divides voxels into eight conditionally independent groups and uses a multiscale architecture modeling voxel occupancy in a coarse-to-fine order. Chen et al. \cite{31} divide the point clouds by progressive downsampling and propose a fully-factorized entropy model which explores the spatial correlation of each level to compress the latent variable. EHEM \cite{32} introduces a grouped context structure to address the serial decoding issue caused by the auto-regression while preserving the compression performance. While all these studies focus on the optimization of the network structure and context information, they overlook the optimization of basic training strategies and the efficiency of context utilization. Specifically, all the models mentioned above, except for CNet \cite{28} and the methods in \cite{16}\cite{30}, use the 255-dimensional one-hot encoding of the eight child nodes of the current node as the label, and the cross-entropy between the one-hot encoding and the probability distribution predicted by the context model as the loss function. As explained in Section. I, this essentially converts the probability distribution prediction problem into a 255-dimensional classification problem. However, this results in two issues. One issue is that the differences in context between different nodes are not significant enough for a context model to accurately predict the probability distribution of node occupancy. The other issue is that the one-hot encoding of node occupancy does not reflect the true probability distribution of node occupancy. 
 \begin{figure}[!t]
\centering
\includegraphics[width=6cm]{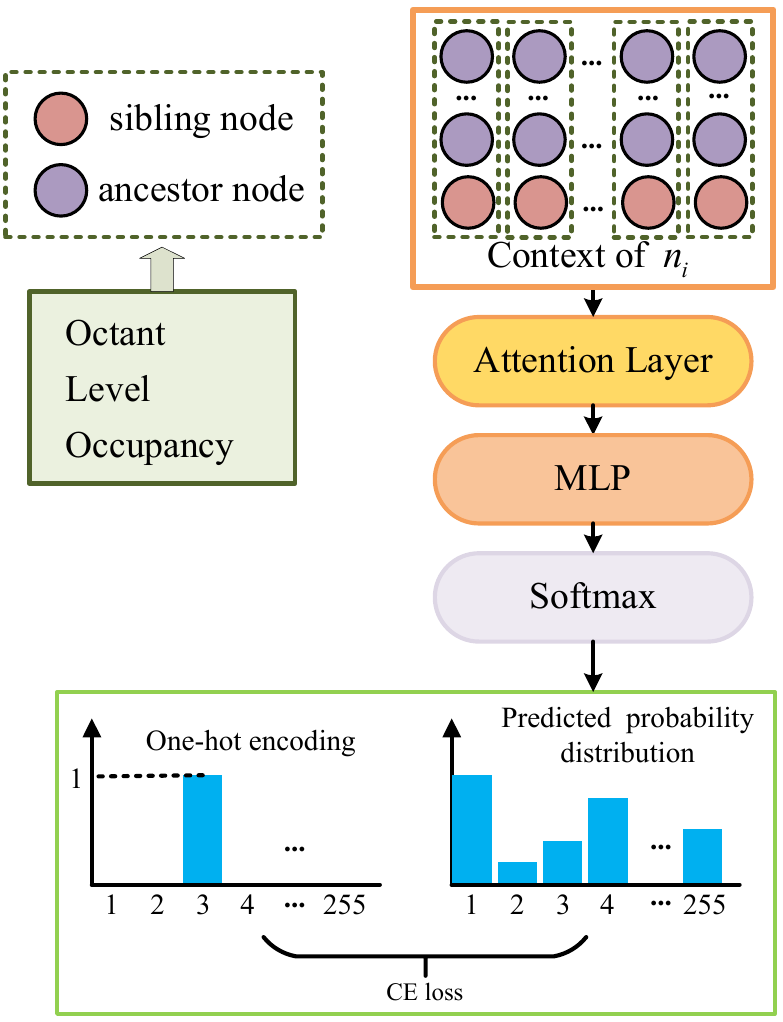}
\caption{Structure of OctAttention.}
\label{fig3}
\end{figure}

 To address these issues, we propose a general structure that can enhance existing context models. The following section will introduce our method using the OctAttention model \cite{14} as an example. For a better understanding, we provide a more detailed introduction to OctAttention (Fig. 3). OctAttention partitions the point cloud using an octree, with each node of the octree being traversed in a breadth-first order. Then, for the current node $n_i$, OctAttention selects the encoded siblings nodes \{$n_{i-N}$, . . . , $n_k$, . . . , $n_{i-1}$\} at the same octree level with a context length of $N$. Moreover, OctAttention introduces $K$ ancestors of the above $N$ nodes in the context to explore the relationship between child nodes and parent nodes. In summary, the contexts consist of ancestor nodes, sibling nodes, and ancestor nodes of sibling nodes. Each node is represented by its occupancy code (1 to 255), level (1 to octree level), and octant (0 to 7). The context is fed into the attention layer \cite{33} to produce a weighted context. Then, the weighted context is sent to a two-layer MLP network. Finally, a Softmax layer generates the probability distribution of the node, which is used for entropy coding. Moreover, it is worth noting that the proposed structure is general and can enhance the performance of multiple learning-based context models for point cloud compression, e.g., VoxelDNN \cite{16} (see Section IV.E).

\begin{figure*}[!t]\centering
  \includegraphics[width=12cm]{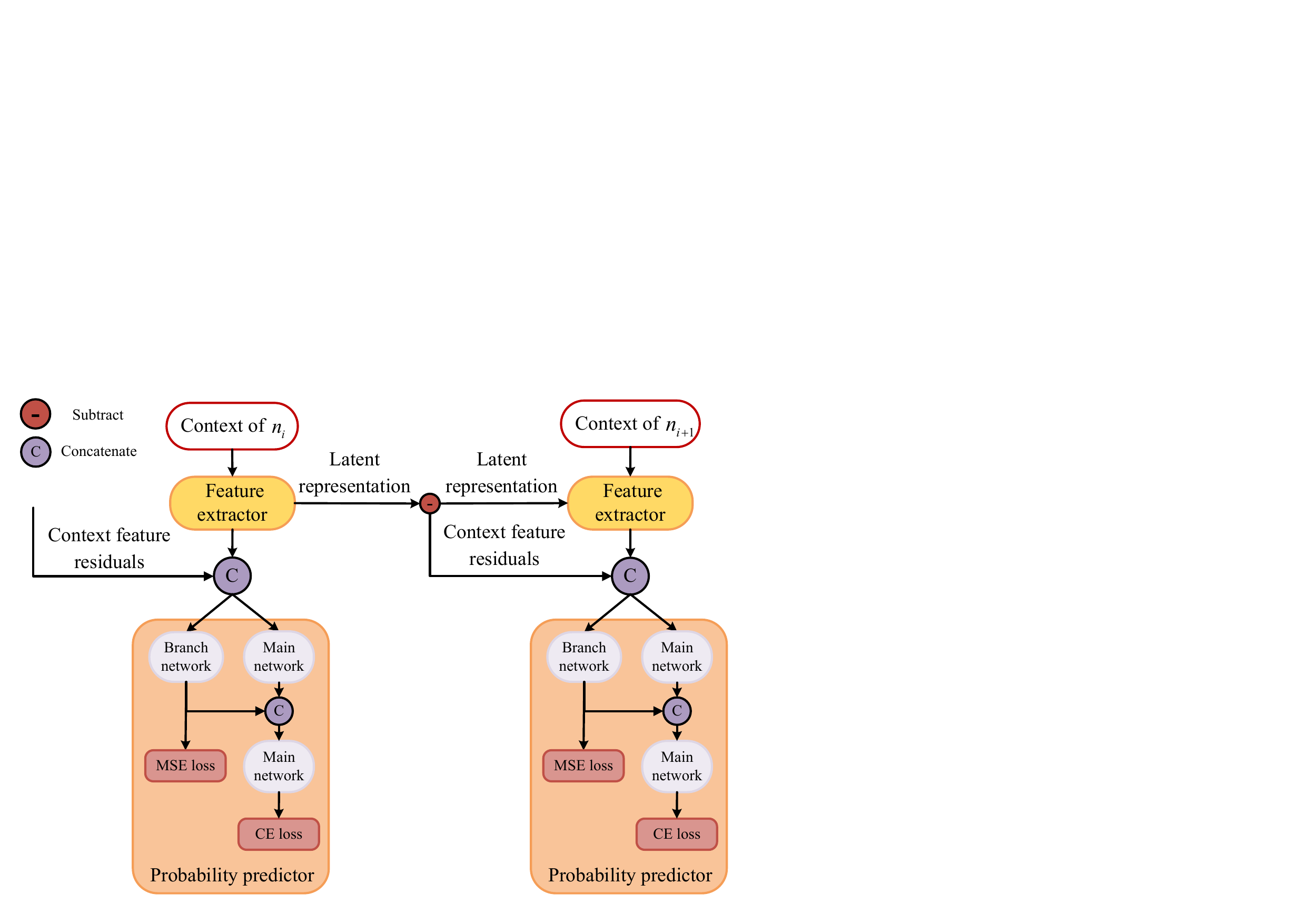}
  \caption{Overall architecture of the proposed structure. The structure consists of a feature extractor, main network, branch network, concatenate module and subtract module. Among them, the feature extractor and main network form the original context model. The subtract module is used to calculate context feature residuals and the concatenate module is used to concatenate the input of the network.}
  \label{fig4}
\end{figure*}
 
\section{Proposed Method}
We propose a method to enhance the context structure for point cloud geometry compression with context feature residuals and multi-loss. In particular, we introduce the context feature residuals into the context models to amplify the difference between neighboring contexts. In addition, we analyze why the cross-entropy loss function is insufficient to accurately measure the difference between predicted and actual probability distributions. We then add an MLP branch that provides accurate labels and loss functions to introduce accurate gradients during the training of the context models.

\subsection{Framework}
Fig. 4 shows the proposed structure. First, a context model is divided into a feature extractor and a probability predictor. The context feature residuals are obtained by calculating the latent representation of the current node $n_{i}$ and the latent representation of the previous node $n_{i-1}$. Then, the latent representation of node $n_i$ and the context feature residuals are concatenated and fed into the probability predictor. For octree-based context models, the main network of the probability predictor usually predicts a 255-dimensional probability distribution, and the branch predicts an 8-dimensional vector to represent the occupancy probabilities of the eight child nodes. For voxel-based models, the main network directly predicts the occupancy probability of a voxel. It should be noted that both context feature residuals and MSE loss-based branches can be introduced into octree-based context models. However, since voxel-based context models directly predict the occupancy probability of a voxel instead of the 255-dimensional probability distribution of the child nodes, only the context feature residuals need to be introduced into voxel-based context models. In the following section, we use our structure to enhance the OctAttention model \cite{14}. We call the resulting model EMR-OctAttention. As shown in Fig. 5, EMR-OctAttention introduces context feature residuals and MSE-based branch compared to OctAttention.

 \begin{figure*}[!t]\centering
  \includegraphics[width=14cm]{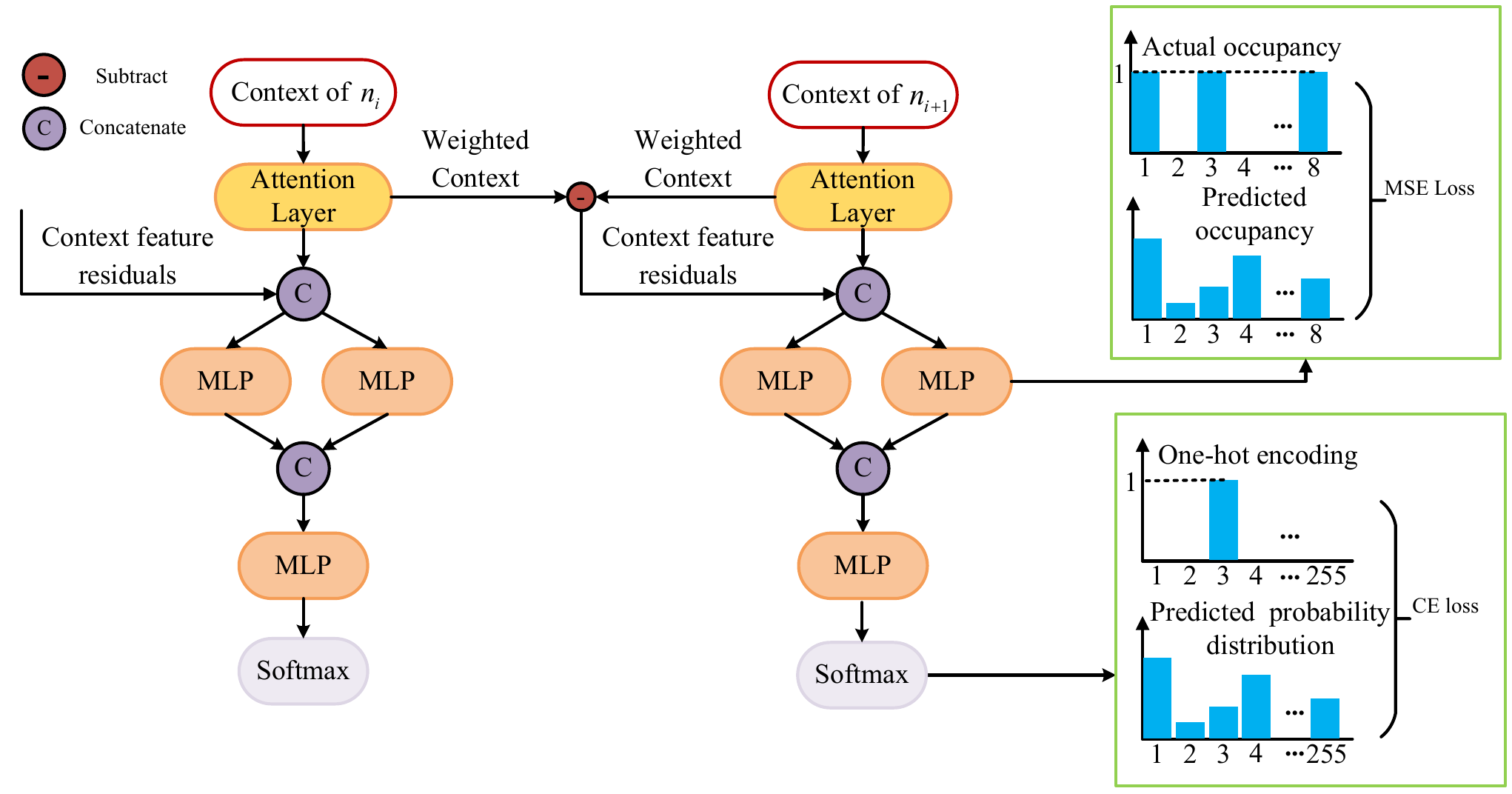}
  \caption{Overall architecture of EMR-OctAttention. The weighted context output from the attention layer is used as a latent representation to calculate the context feature residuals. The context feature residuals are concatenated with the weighted context and fed into two MLPs. One MLP outputs a 255-dimensional probability distribution and is the main network. The cross-entropy between this probability distribution and the one-hot encoding representing the actual occupancy of the node is used as the loss function. The other MLP outputs an 8-dimensional vector representing the occupancy probability of each child node. The mean squared error between this 8-dimensional vector and the actual occupancy of the 8 child nodes is used as the loss function.}
  \label{fig5}
\end{figure*}

\subsection{Context Feature Residuals}
  
OctAttention essentially transforms a prediction of the probability distribution into a 255-dimensional classification problem. However, small differences between different contexts make it challenging for the context model to learn the inter-class differences in context. To amplify the inter-class differences in context, the weighted context is used to calculate the context feature residual of node $x_i$ as
\begin{equation}
\boldsymbol{r_i}=\boldsymbol{wc_i}-\boldsymbol{wc_{i-1}},
\end{equation}
where $\boldsymbol{wc_i}$ is the weighted context of node $x_i$, which is the output of the attention layer. Then, the context feature residuals are concatenated with the weighted context and fed into the MLP. This significantly amplifies the inter-class difference in context. To calculate the inter-class difference, the learning-based model typically uses the mean value of a layer's output as the feature vector of each class. The inter-class difference is then measured by the feature vectors. In this study, we follow the common method in \cite{34} to calculate the inter-class difference. Specifically, all contexts are divided into 255 classes based on node occupancy. The feature vector of each class is the average of the latent representation output of the first layer of the MLP. That is, the feature vector $\boldsymbol{v_j}$ of class $j$ is calculated as
\begin{equation}
\boldsymbol{v_j}=\frac1{n_j}\sum_1^{n_j}H^{(1)}(\boldsymbol{c_k}),
\end{equation}
where $n_j$ is the total number of contexts of class $j$,  $H^{(1)}$ is the first MLP layer in our model, and $\boldsymbol{c_k}$ is the input to $H^{(1)}$. Because the output of the first MLP is a 552-dimensional vector, $\boldsymbol{v_j}$ is a 552-dimensional vector too. Then, the inter-class difference is measured by both the average Euclidean distance and average cosine similarity between all feature vectors. The average Euclidean distance $ad$ is calculated as
\begin{equation}
ad = \frac1{255 \times 255}\sum_{i=1}^{255}\sum_{j=1}^{255}d(\boldsymbol{v_i},\boldsymbol{v_j}),
\end{equation}
where
\begin{equation}
d(\boldsymbol{v_i},\boldsymbol{v_j}) = \sqrt{\sum_{k=1}^{552}(v_{i,k}-v_{j,k})^2} ,
\end{equation}
while  the average cosine similarity $acos$ is calculated as
\begin{equation}
acos = \frac1{255 \times 255}\sum_{i=1}^{255}\sum_{j=1}^{255}\cos(\boldsymbol{v_i},\boldsymbol{v_j}),
\end{equation}
where 
\begin{equation}
\cos (\boldsymbol{v_i},\boldsymbol{v_j}) = \frac{\boldsymbol{v_i} \cdot \boldsymbol{v_j}}{ \Vert \boldsymbol{v_i} \Vert \times \Vert  \boldsymbol{v_j} \Vert}.
\end{equation}

Table 1 shows results for the MPEG 8i dataset. After introducing the context feature residuals, the Euclidean distance between samples increased by 40\%, while the cosine similarity decreased by 13\%. These results show that introducing context feature residuals significantly enhances the inter-class difference, leading the context model to learn the probability distribution more accurately\cite{35,36}.
\begin{table}
\begin{center}
  \caption{Inter-class difference}
  \label{tab1}
  \begin{tabular}{cccc}
    \toprule
     &OctAttention &EMR-OctAttention &Gain\\
    \midrule
    Euclidean distance&10.54 & 14.75& 40\%  \\
    Cosine similarity&0.37 & 0.325 & -13\% \\
  \bottomrule
\end{tabular}
\end{center}
\end{table}
\subsection{MSE-based Branch}
Let $X=\{x_1,x_2,x_3,…,x_i,…\}$ be a sequence of nodes in an octree, according to Shannon's source coding theorem \cite{37}, the lowest bound of its bitrate is its information entropy:

\begin{equation}
  \begin{aligned}
    H(X) &= \mathbb{E}_{X\sim P}[-\log_2p(x_i)]\\
         &= -\sum_{i=1}^{t}  p(x_i)\log_2p(x_i),x_i\in X, \\
  \end{aligned}
\end{equation}
where $p(x_i)$ is the occupancy probability of node $x_i$, and $t$ is the number of nodes in the octree. The ground truth probability distribution $P(X)$ can be expressed as
\begin{equation}
P(X)=\prod_{i=1}^{t} p(x_i|x_{i-1},x_{i-2},...,x_1).
\end{equation}
However, since $P(X)$ is unknown, the context model $\boldsymbol{\omega}$ estimates it as
\begin{equation}
Q(X)=\prod_{i=1}^{t} q(x_i|x_{i-1},x_{i-2},...,x_1;\boldsymbol{\omega}),
\end{equation}
where $q(x_i)$ is the occupancy probability predicted by the context model $\boldsymbol{\omega}$. Ideally, the loss function of a context model should be defined as the cross-entropy between $P(X)$ and $Q(X)$:
\begin{equation}
Loss=-\sum_{i=1}^{t}  p(x_i)\log_2q(x_i),x_i\in X.
\end{equation}

Most entropy models replace $P(X)$ with the one-hot encoding of nodes and use the one-hot encoding as the label (Fig. 2). However, the one-hot encoding of node occupancy does not reflect the true probability distribution of node occupancy. Moreover, most entropy models use the cross-entropy between one-hot encoding and $Q(X)$ as the loss function, which essentially treats probability distribution prediction as a classification task. unlike traditional classification tasks where the differences between two classes are significant, the differences between any two classes are unbalanced in OctAttention. For example, only one node is different between the 255th occupancy configuration ‘11111111' and the 254th occupancy configuration ‘11111110’, while seven nodes are different between the 255th and 1st. As shown in Fig. 2, although the cross-entropy between probability distribution A and training label C, and that between probability distribution B and training label C are identical, probability distribution A is closer to label C. This is because probability distribution A assigns a higher probability to classes that are similar to the label. In conclusion, using cross-entropy as the loss function cannot accurately calculate the difference between one-hot encoding and the predicted probability distribution.

The reason for the above problem is that the probability distribution output by OctAttention includes both the number and position of occupied child nodes. For example, the 254th occupancy configuration represents ‘11111110’, where seven child nodes are occupied and their positions are the first seven child nodes in the scanning order. While determining the position of occupied child nodes is a classification problem, finding the number of occupied child nodes is a regression problem. OctAttention combines this classification problem and this regression problem into a 255-dimensional classification problem. As a result, the cross-entropy cannot accurately calculate the difference between the one-hot encoding and the predicted probability distribution.

Based on the above analysis, one possible approach to guide the learning of context models is to use more accurate labels. However, it is difficult to obtain more accurate labels than one-hot encoding as a point cloud is not a memoryless source.An alternative approach is to separate the regression task and use the regression result as a feature to assist the learning of the main network. As shown in Fig. 5, we add an MLP branch after the attention layer. The last layer of this branch outputs an 8-dimensional vector, where each dimension represents the occupancy of a child node. During the training phase, the MSE loss is used to calculate the loss between the output of this branch and the actual occupancy of each child node. This converts the number of occupied child nodes into eight regression tasks for processing. As the output 8-dimensional vector of this branch represents the probability of each child node being occupied, it will contain information on the number of occupied child nodes, which is useful for the training of the main network. Finally, the output of this branch is concatenated with the output of the main network at the same layer and then fed into the final layer of the main network to get a more accurate prediction of the probability distribution.
\subsection{Loss Function}
The main network is trained with the cross-entropy loss between the one-hot encoding of node $x_i$ and the probability distribution output by the main network:

\begin{equation}
L_{CE}=-\sum_{i=1}^{t} \boldsymbol{e}(x_i)\log_2q(x_i|\boldsymbol{c_i};\boldsymbol{\omega}),x_i\in X,
\end{equation}
where $t$ is the number of nodes, $\boldsymbol{c_i}$ is the context of node $x_i$ and $\boldsymbol{e}(x_i)$ is the one-hot encoding. The branch network is trained with the MSE loss between the actual occupancy  $\boldsymbol{l_i}$ of node $x_i$ and the 8-dimensional vector $\boldsymbol{o_i}$ output by the branch network:
\begin{equation}
L_{MSE}=-\frac1{t}\sum_{i=1}^{t}{\|  \boldsymbol{l_i}- \boldsymbol{o_i} \|_2}. 
\end{equation}

\begin{table*}
\begin{center}
  \caption{Average bitrate (in BPIP) of EMR-OctAttention on MVUB and MPEG 8i}
  \label{tab2}
  \begin{tabular}{ccccc}
    \toprule
    Dataset&	Point cloud sequence& EMR-OctAttention	&Gain over G-PCC	&Gain over OctAttention\\
    \midrule
\multirow{7}*{8i} & Loot	& 0.588 	&-20.4\%&-2.52\% \\
~ &RedandBlack &	 	0.700 &	-15.8\%	&-2.41\% \\
~ & Thaidancer10&0.618 &	-12.8\%	&-2.11\% \\
~&Thaidancer9 &	0.610 &	-16.1\%	&-3.65\%\\
~ &Boxer10	 &	0.554 &	-15.8\%	&-2.48\% \\
~&Boxer9 &	0.558 	&-21.5\%&	-2.96\%\\
~&Average&	0.605 &	-17.1\%	&-2.68\% \\
    \bottomrule
    \multirow{5}*{MVUB} & Phil10	& 	0.761 &	-19.9\%	&-2.65\% \\
~&Phil9	 &	0.800 	&-19.9\%&	-2.59\%\\
~&Ricardo10	 	&0.688 	&-22.7\%	&-1.94\% \\
~&Ricardo9		&0.686 	&-24.6\%	&-2.60\%\\
~&Average		&0.734	&-21.8\%	&-2.46\%\\
    \bottomrule
  \end{tabular}
\end{center}
\end{table*}

\begin{table*}
\begin{center}
  \caption{Average bitrate (in BPIP) and reconstruction quality of EMR-OctAttention on SemanticKITTI}
  \label{tab3}
  \begin{tabular}{cccccc}
    \toprule
    Octree Level&	CD&	D1-PSNR	&	EMR-OctAttention&	Gain over G-PCC&	Gain over OctAttention\\
    \midrule

12&	2.35E-04&	77.02&	3.56	&-10.0\%&	-1.44\% \\
11&	4.69E-04	&70.99&	1.92	&-20.1\%	&-1.54\% \\
10&	9.39E-04&	64.97&	0.886&	-29.3\%&	-2.68\%\\
9&	1.88E-03&	58.94&	0.360&	-25.9\%&	-2.17\% \\
8&	3.77E-03	& 52.91&	0.131&	-17.5\%&	-2.38\% \\
Average&	1.46E-03 &	64.97& 		1.37& 	-20.6\%	&-2.04\%\\

    \bottomrule
  \end{tabular}
\end{center}
\end{table*}

\section{Experiments}
To validate our method, we used it to improve OctAttention \cite{14}. We trained two EMR-OctAttention models: one for geometry compression of object point clouds and the other for geometry compression of LiDAR point clouds. While the training datasets and epoch for the two models differ, the models’ structures and hyperparameters are identical.

To show that context feature residuals are also effective for voxel-based models, we used our method to improve the VoxelDNN model \cite{16}. We call the enhanced VoxelDNN model ER-VoxelDNN. Since VoxelDNN is not suitable for compressing LiDAR point clouds, we evaluated ER-VoxelDNN on object point clouds only.

\subsection{Datasets}
For object point clouds, we used Microsoft Voxelized Upper Body (MVUB) \cite{38} and 8i Voxelized Full Bodies (MPEG 8i) \cite{39}. MVUB contains voxelized point cloud sequences that record the upper bodies of individuals, with a geometry accuracy of either 10 or 9 bits. The MPEG 8i dataset uses point cloud sequences with a geometry accuracy of either 10 or 12 bits to record the complete human body. Following the settings in \cite{14}, we used the andrew10, david10, and sara10 point cloud sequences from MVUB, and the soldier10 and longdress10 sequences from MPEG 8i to create a training set. Other point cloud sequences were used for testing. As Boxer12 and Thaidancer12 in MPEG 8i have a geometry accuracy of 12 bits, we downsampled them to 10 and 9 bits, respectively, for testing purposes.

For LiDAR point clouds, we used SemanticKITTI \cite{40}, a large sparse LiDAR dataset for autonomous driving consisting of 43552 scans and 454.9 million points. SemanticKITTI is a point cloud dataset collected from various environments such as urban areas, countryside, and highways. Following the settings in \cite{14}, we normalized the raw data to [-1, 1] and used sequences 00 to 10 for training and 11 to 21 for testing.

\subsection{Experimental Details}
\noindent {\bf{Implementation settings:}} We implemented EMR-OctAttention in PyTorch. We followed the hyperparameter settings in \cite{14}, using a batch size of 32 and a context size of 1024. We set the learning rate of the Adam optimizer to 0.001 with a learning rate decay $\Gamma=0.95$. For LiDAR point clouds, we set the epoch to 8. Since object point clouds have a faster training speed than LiDAR point clouds, we set the epoch to 80. We first fixed the MLPs of the main network and trained the branch network for 2 and 20 epochs on LiDAR point clouds and object point clouds, respectively. Then we fixed the branch network and trained the  main network for 6 and 60 epochs on LiDAR point clouds and object point clouds, respectively. We always choose $N=1024$ adjacent nodes and $K=4$ ancestor nodes, which has been proven to be effective in OctAttention.

We implemented ER-VoxelDNN in Tensorflow. Following the hyperparameter settings in \cite{33}, we trained ER-VoxelDNN with the Adam optimizer, a learning rate of 0.001, and a batch size of 8 for 50 epochs.

Both training and testing of EMR-OctAttention and ER-VoxelDNN were conducted on an Intel(R) Xeon(R) Gold 6148 CPU and a GeForce RTX 4090 GPU with 24 GB memory. 

\noindent {\bf{Baseline:}} We compared EMR-OctAttention and ER-VoxelDNN with  OctAttention \cite{14} and VoxelDNN \cite{33} for the same implementation settings. We also compared them with G-PCC test model TMC13 v23.0 \cite{17}.

\noindent {\bf{Evaluation Metrics:}} For object point cloud compression, a 10-level octree was used to fully partition an object point cloud, and the encoding was lossless. Consequently, the performance was fully characterized by the bitrate, which we measured in bits per input point (BPIP).

As LiDAR point clouds are sparse, fully partitioning them with an octree is inefficient. Therefore, LiDAR point clouds were partitioned into 12 levels for model training. For testing, the point clouds were partitioned into 8-12 levels to evaluate our model at different bitrates. The incomplete partition of LiDAR point clouds leads to distortion. Point-to-point PSNR (D1 PSNR) \cite{41} and Chamfer distance (CD) were used to measure the geometry reconstruction quality. The geometry reconstruction quality of the decompressed LiDAR point cloud only depends on the number of octree levels in the point cloud partition. For the same number of octree levels, the reconstructed point clouds of our method and other octree-based lossless compression methods is the same.
\begin{figure}[!t]\centering
  \includegraphics[width=8cm]{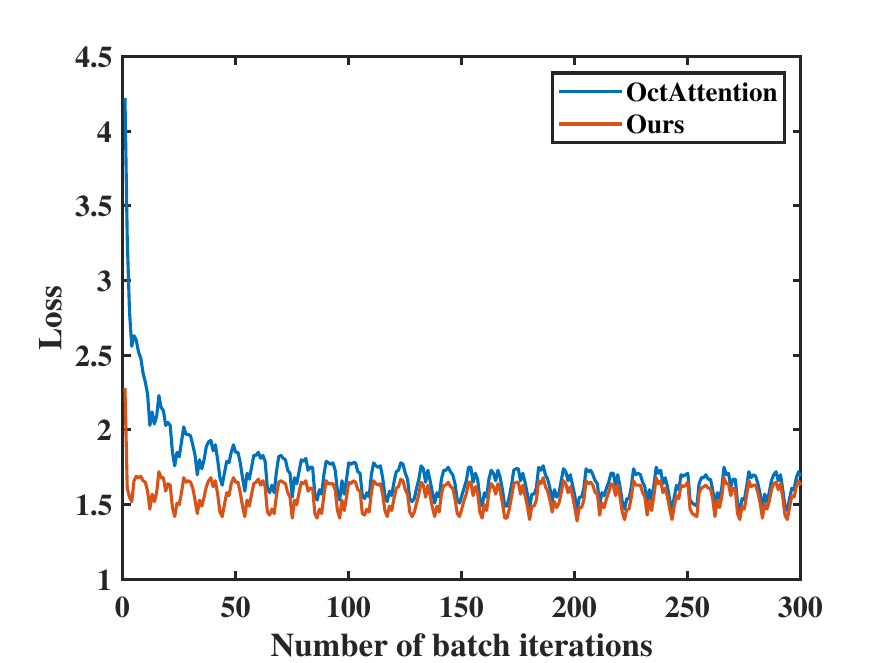}
  \caption{Loss vs. Number of batch iterations for EMR-OctAttention and OctAttention}

  \label{fig6}
\end{figure}

\begin{table*}
\begin{center}
  \caption{Ablation experiment on object point clouds}
  \label{tab4}
  \begin{tabular}{ccccccccc}
    \toprule
    Dataset&	Point cloud sequence&OctAttention&\multicolumn{2}{c}{EM-OctAttention} &	\multicolumn{2}{c}{ER-OctAttention}&\multicolumn{2}{c}{EMR-OctAttention}\\
    &&BPIP	&BPIP 	&Gain&	BPIP &	Gain&	BPIP 	&Gain\\
    \midrule

\multirow{7}*{8i} & Loot	&0.603& 	0.601 &	-0.35\%&	0.592& 	-1.86\%	&0.588 &	-2.52\% \\
~ &RedandBlack &	0.717 	&0.716 &	-0.22\%	&0.702 &	-2.06\%	&0.700 	&-2.41\% \\
 & Thaidancer10&	0.631& 	0.631& 	-0.02\%&	0.621& 	-1.62\%&	0.618& 	-2.11\% \\
~&Thaidancer9 &	0.633& 	0.633& 	0.05\%	&0.615& 	-2.89\%	&0.610 	&-3.65\% \\
~ &Boxer10	&0.568& 	0.564& 	-0.65\%&	0.558& 	-1.83\%&	0.554 &	-2.48\% \\
~&Boxer9&	0.575 &	0.575 &	-0.05\%&	0.563 &	-2.17\% &	0.558 	&-2.96\%\\
~&Average&	0.621 &	0.620 &	-0.21\%	&0.608 	&-2.07\%	 &0.605 	&-2.68\%\\
    \bottomrule
    \multirow{5}*{MVUB} & Phil10	&0.782 &	0.773 &	-0.86\%	&0.763 	&-2.39\%	&0.761 	&-2.65\% \\
~&Phil9	&0.821 &	0.817 &	-0.50\%	&0.803 	&-2.29\% &	0.800 &	-2.59\%\\
~&Ricardo10	&0.701 	&0.698 	&-0.46\%	&0.690	&-1.65\%	 &0.688 	&-1.94\% \\
~&Ricardo9	&0.705	&0.701 	&-0.45\%	&0.691 	&-1.96\%	 &0.686 &	-2.60\%\\
~&Average	&0.752 	&0.747 	&-0.57\%	&0.736 &	-2.06\%	&0.734 &	-2.46\%\\
    \bottomrule
  \end{tabular}
\end{center}
\end{table*}

\subsection{EMR-OctAttention}
As shown in Table 2, EMR-OctAttention yielded the lowest bitrate on MVUB and MPEG 8i. On average, EMR-OctAttention reduced the bitrate by 17.1\% on MPEG 8i and 21.8\% on MVUB compared to G-PCC. Furthermore, EMR-OctAttention outperformed OctAttention by reducing the bitrate by 2.68\% on MPEG 8i and nearly 2.46\% on MVUB. 

As shown in Table 3, EMR-OctAttention achieved the lowest bitrate on SemanticKITTI. On average, EMR-OctAttention reduced the bitrate by 20.6\% compared to G-PCC, and 2.04\% compared to OctAttention. The experimental results demonstrate that EMR-OctAttention effectively reduced the bitrate in point cloud geometry encoding for LiDAR point clouds. 

Fig. 6 shows the loss of EMR-OctAttention and OctAttention during training on the object point cloud datasets. EMR-OctAttention achieved significantly lower loss at the beginning of training compared with OctAttention. This can be attributed to its ability to enhance the inter-class differences in context. As a result, EMR-OctAttention quickly learned the inter-class differences in context. As training progressed, the performance of OctAttention gradually approached that of EMR-OctAttention. This is because OctAttention can also distinguish inter-class differences in context with more training.

\begin{table*}
\begin{center}
  \caption{Ablation experiment on LiDAR point clouds}
  \label{tab5}
  \begin{tabular}{cccccccc}
    \toprule
    Octree Level&OctAttention&\multicolumn{2}{c}{EM-OctAttention} &	\multicolumn{2}{c}{ER-OctAttention}&\multicolumn{2}{c}{EMR-OctAttention}\\
    &BPIP	&BPIP 	&Gain&	BPIP &	Gain&	BPIP 	&Gain\\
    \midrule

12	&3.61 	&3.53 	&-2.22\% &	3.54 &	-1.83\%	&3.56	&-1.44\% \\
11	&1.95 	&1.91 	&-2.15\%	 &1.91 	&-1.90\% &	1.92	&-1.54\%\\
10	&0.910 	&0.880 	&-3.33\% &	0.882 	&-3.12\%	 &0.886 	&-2.68\% \\
9&	0.368& 	0.354 &	-3.83\%& 	0.353& 	-4.08\%& 	0.360& 	-2.17\%\\
8	&0.134 	&0.129 	&-3.95\%	 &0.130 	&-3.20\%	 &0.131& 	-2.38\%\\
Average	&1.394	&1.360	&-3.10\%	 &1.364	&-2.82\%	 &1.371 &	-2.04\%\\

    \bottomrule
  \end{tabular}
\end{center}
\end{table*}

\subsection{Ablation Study and Analysis}
In this section, we assess the effectiveness of the context feature residuals and the MSE Loss based branch. We trained two models: the first (ER-OctAttention) only adds the proposed context feature residuals, while the second (EM-OctAttention) only adds the MSE Loss based branch. Both ER-OctAttention and EM-OctAttention use the training and testing methods of EMR-OctAttention. 

\begin{figure}[!t]\centering
  \includegraphics[width=6cm]{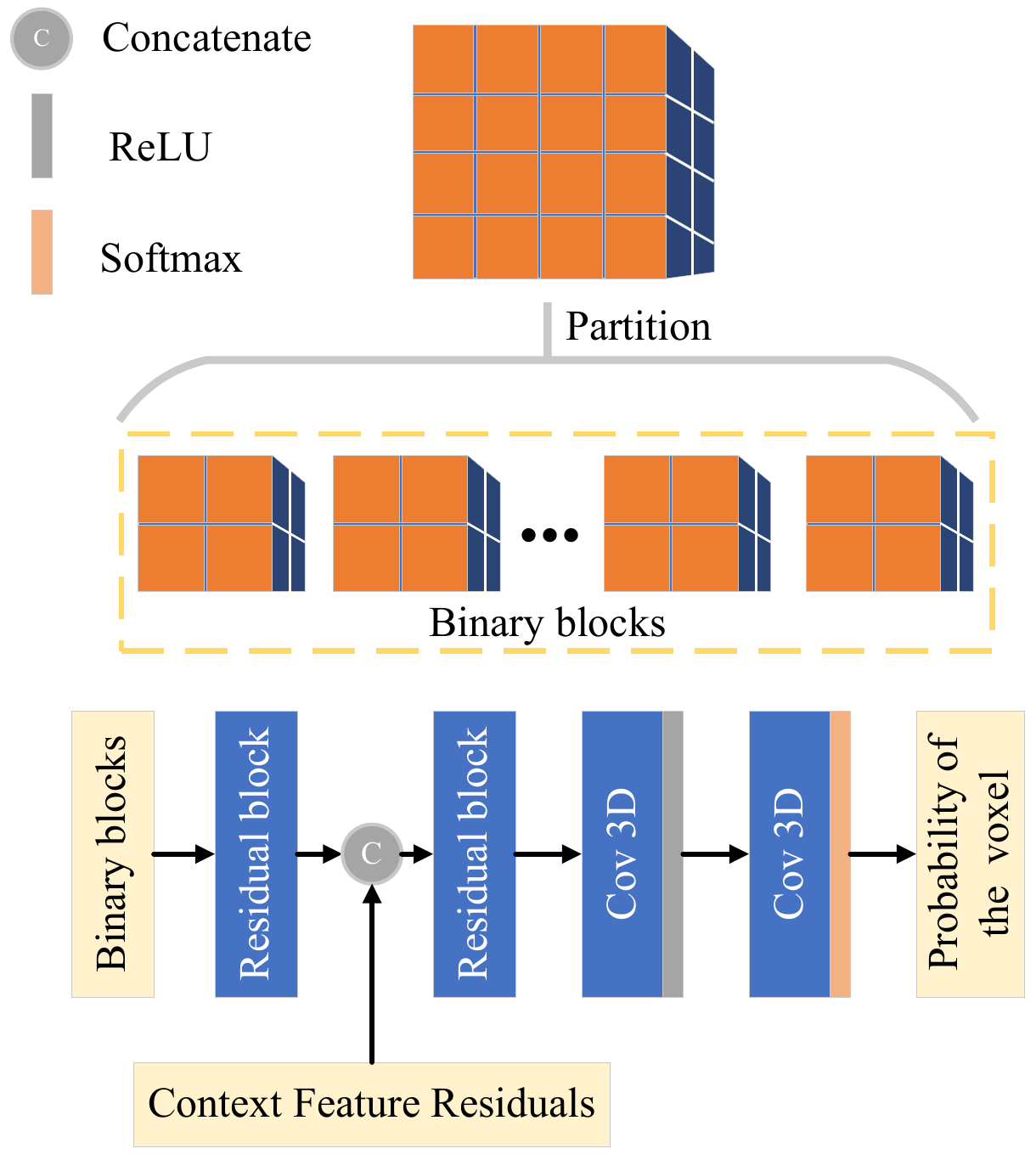}
  \caption{Overall architecture of ER-VoxDNN.}

  \label{fig7}
\end{figure}

Table 4 presents the experimental results for object point cloud compression. All three proposed models show improvement for point cloud sequences in MPEG 8i and MVUB. Among them, EMR-OctAttention achieved the lowest BPIP and reduced the average bitrate by 2.68\% on MPEG 8i and 2.46\% on MVUB.  ER-OctAttention reduced the average bitrate by 2.07\% on MPEG 8i and 2.06\% on MVUB. However, EM-OctAttention did not show a significant improvement in comparison to OctAttention, and it only reduced the average bitrate by 0.21\% on MPEG 8i and 0.57\% on MVUB. This is because in object point clouds, the number of occupied child nodes of the current node is relatively stable, which makes the influence of EM-OctAttention on predicting the number of occupied child nodes is small. 

\begin{table*}
\begin{center}
  \caption{Average bitrate (in BPIP) on MVUB and MPEG 8i}
  \label{tab6}
  \begin{tabular}{ccccc}
    \toprule
    Dataset&	Point cloud sequence& VoxelDNN &	ER-VoxelDNN	&Gain over VoxelDNN\\
    \midrule
\multirow{5}*{8i} & Loot	&0.653 &0.649 &-1.55\% \\
~ &RedandBlack &	0.721&	0.709 		&-1.67\% \\
~ & Thaidancer10&	0.667	&0.650   &-1.93\% \\
~ &Boxer10	&0.592	&0.583 &	-1.59\% \\
~ & Thaidancer9&	0.779	&0.769   &-1.28\% \\
~ &Boxer9	&0.721	&0.705 &	-2.22\% \\
~&Average&	0.688&	0.676 &-1.71\% \\
    \bottomrule
    \multirow{3}*{MVUB} & Phil10	&0.771	&0.763 &-1.10\% \\
~&Ricardo10	&0.712	&0.704 	&-1.14\% \\
~&Phil9	&0.835	&0.829 	&-0.72\% \\
~&Ricardo9	&0.739	&0.734 	&-0.67\% \\
~&Average	&0.764 	&0.757 	&-0.91\%\\
    \bottomrule
  \end{tabular}
\end{center}
\end{table*}

\begin{figure}[!t]\centering
  \includegraphics[width=8cm]{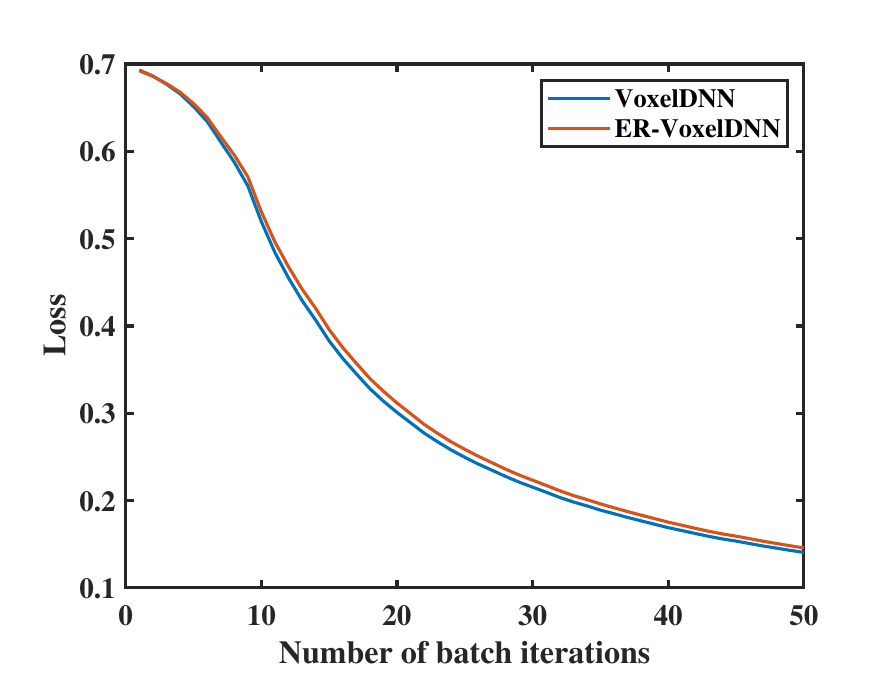}
  \caption{Loss vs. Number of batch iterations for ER-VoxelDNN and VoxelDNN}

  \label{fig8}
\end{figure}

Table 5 shows the results for LiDAR point cloud compression. All three proposed models decreased the BPIP. In contrast to the results for object point clouds, EM-OctAttention showed better performance than ER-OctAttention and reduced the average bitrate by 3.10\%. Due to the sparsity of LiDAR point clouds, adjacent nodes in the encoding sequence end up being far apart, resulting in greater differences in context. Consequently, the context feature residuals are less effective. 

\begin{table*}
\begin{center}
  \caption{Computational complexity comparison}
  \label{tab7}
  \begin{tabular}{cccccccc}
    \toprule
    &OctAttention&\multicolumn{2}{c}{EM-OctAttention} &	\multicolumn{2}{c}{ER-OctAttention}&\multicolumn{2}{c}{EMR-OctAttention}\\
    &result	&result 	&Gain&	result &	Gain&	result 	&Gain\\
    \midrule
  Model size (M)	  &28.0	 &29.3 &4.6\%  &29.2 &4.3\%&31.7 &13.2\% \\
Encoding time (s)	&0.263	&0.274	&4.2\%&0.272	&3.4\% &0.296	&12.5\%\\
Decoding time (s)	&345	&362	&4.9\% &358	&3.8\%&392	&13.6\%\\

    \bottomrule
  \end{tabular}
\end{center}
\end{table*}

\subsection{ER-VoxDNN}

In this section, to verify the generality of our method, we use it to enhance the voxel-based context model VoxelDNN. VoxelDNN first coarsely partitions an n-depth point cloud up to level ${n-6}$. This gives 64 non-empty binary blocks of size ${2^n \times 2^n \times 2^n}$ voxels. Each block is then encoded separately. VoxelDNN encodes each voxel within a block in raster scan order. All the encoded voxels in the current binary block form the context and serve as input to VoxelDNN, enabling the direct prediction of the occupancy probability for the current voxel. This is a regression problem. Therefore, ER-VoxelDNN does not need to introduce the MSE loss to solve the problem of inaccurate cross-entropy loss based on one-hot encoding. Since the contexts of two adjacent nodes differ by only one node, we only need to introduce the context feature residuals. We call the enhanced model ER-VoxelDNN. Specifically, the features output by the first residual block are used as context features, and the context feature residuals are calculated. Then, the context feature residuals are concatenated with the context features and input into subsequent networks. The structure of ER-VoxelDNN is shown in Fig. 7.

The experimental results are shown in Table 6. ER-VoxelDNN reduced the average bitrate by 1.71\% on MPEG 8i and 0.91\% on MVUB. The experimental results demonstrate that introducing context feature residuals is effective in voxel-based models.

Fig. 8 shows the loss of ER-VoxelDNN and VoxelDNN during training on the object point cloud datasets. The loss for ER-VoxelDNN between the 10th and 50th batches during training was significantly lower than for VoxelDNN. As in Fig. 6, this is because the context feature residuals help ER-VoxelDNN to learn the inter-class differences in context more quickly. Compared with octree-based models in Fig. 6, the loss of ER-VoxelDNN was smaller and the loss curve was smoother. The voxel-based context model only needs to predict the occupancy probability of a single voxel, which is significantly simpler compared to predicting the occupancy probability distribution for eight child nodes in the octree-based model. Therefore, the voxel-based model had a smaller loss and a more stable transformation of the loss between different batches.

\subsection{Complexity Comparison}
We computed the average encoding time and the average decoding time of each frame in the loot and redandblack point cloud sequences from the MPEG 8i dataset. Table 7 and Table 8 show the computational complexity of our models. For EMR-OctAttention, the context feature residuals increase the input dimension of the MLPs. On the other hand, the MLP branch increases the number of parameters. As a result, EMR-OctAttention was 13.2\% larger than OctAttention. EMR-OctAttention increased the encoding time by 12.5\% and decoding time by 13.6\% compared to OctAttention. 
Table 8 shows the time complexity of ER-VoxelDNN. ER-VoxelDNN's context feature residuals led to a 3.16\% increase in encoding time and a 4.06\% increase in decoding time. Since the convolution kernels in VoxelDNN are shared,  the context feature residuals increase the input dimension but do not increase the number of parameters in the model.

\begin{table}
\begin{center}
  \caption{Computational complexity comparison}
  \label{tab8}
  \begin{tabular}{cccc}
    \toprule
     & VoxelDNN 	&ER-VoxelDNN	&Increase \\
    \midrule
  Model size (M)	  &3.4	&3.4 &0\% \\
Encoding time (s)	&2248	&2319	&3.16\%\\
Decoding time (s)	&5686	&5917	&4.06\%\\
  \bottomrule
\end{tabular}
\end{center}
\end{table}

\section{Conclusions and Future Work}
We proposed a general method to enhance context models for geometry point cloud compression. Our method is designed to address the problem of small context differences and the inaccuracy of the loss function based on cross-entropy and one-hot encoding by using multi-loss and context feature residuals. Our method incorporates context feature residuals in training to enhance the differences between contexts. It also adds another MLP network branch that uses the MSE between its output and node occupancy as the loss function to ensure accurate gradients in backpropagation. We showed that our method can improve the performance of an octree-based model (OctAttention) and a voxel-based model (VoxelDNN) on the object point cloud datasets MPEG 8i and MVUB, as well as the LiDAR point cloud dataset SemanticKITTI.

\end{document}